\documentclass[prb,twocolumn]{revtex4}
\usepackage{amssymb}
\usepackage{graphicx}
\usepackage{amsfonts}
\usepackage{amsmath}

\begin{document}

\title{Quantum pumping in deformable quantum dots}

\author{F. Romeo$^{1}$, and R. Citro$^{1,2}$}
\affiliation{$^{1}$Dipartimento di Fisica ''E. R. Caianiello'' and
C.N.I.S.M., Universit{\`a} degli Studi di Salerno, Via S. Allende,
I-84081 Baronissi (Sa), Italy\\
$^{2}$Laboratorio Regionale INFM-CNR SuperMat, Via S. Allende,
I-84081 Baronissi (Sa), Italy }

\begin{abstract}
The charge current pumped adiabatically through a deformable
quantum dot is studied within the Green's function approach.
Differently from the non-deformable case, the current shows an
undefined parity with respect to the pumping phase $\varphi$. The
unconventional current-phase relation, analyzed in the weak
pumping regime, is due to a dynamical phase shift $\phi_D$ caused
by the elastic deformations of the central region (classical
phonons). The role of the quality factor $Q$ of the oscillator,
the effects induced by a mechanical resonance and the implications
for current experiments on molecular systems are also discussed.
\end{abstract}



 \maketitle

\section{ Introduction}
The idea of quantum pumping, i.e. of producing a dc current at
zero bias voltage by time periodic modulation of two system
parameters, dates back to the work of Thouless \cite{thouless}.
The pumping is {\it adiabatic} when the parameters change slowly
as compared to all internal time scales of the system and the
average charge pumped per period does not depend on the specific
time dependence of the parameters. Using the concept of emissivity
proposed by B\"{u}ttiker et al.\cite{buttiker_injectivity},
Brouwer\cite{brouwer98} related the charge pumped in a period to
the derivatives of the instantaneous scattering matrix of the
conductor with respect to the time-varying parameters. Since then,
a general framework to compute the pumped charge through a
conductor has been developed for noninteracting and interacting
electrons\cite{various,various-interacting}. Then the interest in the pumping phenomenon has shifted to the
experimental\cite{exp-dot} investigations of confined
nanostructures, as quantum dots, where the realization of the
periodic time-dependent potential can be achieved by modulating
gate voltages applied to the structure\cite{exp-dot}.

On the other hand, the modern miniaturization techniques have
allowed the realization of artificial structures of the same
length scale of molecular objects so that the charge distribution
and shape of the system may well change during their operation.
The process of hybridization obtained by joining artificial
manmade structures with macromolecules makes possible to get
systems whose electrical response is strongly affected by
electromechanical tunable interaction. This happens because at
molecular scale the elastic forces controlling the structure of
the system are of the same order of magnitude with respect to the
electrostatic forces related to the charge distributions. The
interplay between electrostatic and structural degrees of freedom
can be experimentally studied by means of the carbon-based
technology allowing a full integration between manmade
nanostructures and molecular conductors (as in particular the
fullerene $C_{60}$). In such systems, whose theory has been
formulated by Gorelik \textit{et al.} \cite{gorelik-shuttle}, the
interplay between mechanical and electromagnetic degrees of
freedom is responsible for a mechanical instability that causes a
self-oscillating behavior due to a self-consistent bistable
potential. The current induced by this so-called \textit{charge
shuttle phenomenon}\cite{donarini-phd} is proportional to the
frequency of the oscillating part of the system, while the
electric charge transported per cycle is an integer multiple of
the elementary charge $-e$. The coupling between charge transport
and vibrational properties of these molecular systems has recently
been observed by H. Park
 \textit{et al.} \cite{parkc60} in an experiment on a  single-molecule
$C_{60}$-based transistor. Similar effects have also been observed in experiments on the
transport properties of suspended carbon nanotubes where the
activation of the breathing modes (i.e. radial deformations of the
nanotube) has been obtained using an STM (Scanning tunneling microscope)
tip \cite{leroy-nanotube}. These systems are nowadays known as
nanoelectromechanical systems (NEMS) and are very important in
clarifying the role of the electron-phonon interactions in
molecular conductors and manmade
artificial systems.\\
Within the previous scenarios, an interesting question is what
type of effect would induce the mechanical degrees of freedom
(oscillations) on quantum pumping properties. In particular, one
can imagine to realize an adiabatic quantum pump with a deformable
quantum dot capable of changing its configuration under the
combined effect of electrostatic and elastic forces.
 This proposal can be realized, for instance, by coupling a
suspended nanotube\cite{telescoping-cnt} to external leads through
two tunable tunnel barriers. Suspended nanotubes of $\sim 0.5
\mu$m of length present a mechanical resonance frequency of $\sim
200$MHz, i.e. of the same order of the adiabatic pumping frequency
and a quality factor $Q\sim 100-150$ (see for instance Fig.3 of
Ref.[\onlinecite{telescoping-cnt}]). Thus the mentioned molecular
systems are good candidates to explore the adiabatic quantum
pumping through a deformable system.\\
In this work, by employing a Green's function approach, we derive
an expression of the adiabatic pumping current through a
deformable quantum dot subject to a static electric field $E_g$
generated by a back gate and coupled to two external leads kept at
the same chemical potential. The pumped current in this system is
generated by the adiabatic variation of the tunneling rates
between the noninteracting leads and the deformable quantum dot.
During the pumping cycle the deformable dot is charged with a
finite density and reacting to the presence of the electric field
$E_g$ by deforming itself and changing its energy. The dynamical
effects related to the mechanical deformation of the system are
considered as adiabatic and are described by using a classical
formalism. The small deformation limit is considered to avoid
interesting but very complicated
non-linear effects\cite{cleland-book}.\\
An important point is that the parametric pump we have in mind is
very different from a shuttle device. Indeed, the shuttle consists
of a self-oscillating system able to self-sustain a mechanical
oscillation of a small grain when a dc bias is applied to the
external leads. In this setup the charge is transported on the
shuttle and is released to the closest lead, the charge on the
grain being almost constant during the shuttling time. Thus a
crucial ingredient is the possibility to reach closely a given
lead to enhance the tunneling rate which then depends strongly on
the position of the shuttle in the direction of charge transport
(longitudinal direction). Alternatively, we are proposing here a
parametric pumping, the only difference from the original work of
Thouless\cite{thouless} being the inclusion of a classical
variable associated to the transverse oscillations of the central
region subjected to a back gate. In our pumping setup the
tunneling rates are unsensitive to the position of the central
region in the longitudinal direction and only depends on the
transverse motion of the center of mass of the central region. The
self-oscillating phenomena present in the shuttle dynamics (see
e.g. [\onlinecite{fazio-pistolesiprl05}]) are then absent in our
proposal.

The plan of the paper is the following. In Sec.\ref{sec:model} we
introduce the model Hamiltonian and the relevant parameters. In
Sec.\ref{sec:x-dyn} the classical dynamics of the variable $x$
describing the motion of the center of mass of the central region
is studied within the small deformation limit (i.e. in linear
response). In Sec.\ref{sec:weak pump} we derive the weak pumping
formula for the current which includes the effects of the
additional phase shift $\phi_D$ due to the dynamics of the
$x$-variable. In Sec.\ref{sec:results} we present the results of
the pumped current at zero temperature by discussing the role of
the mechanical oscillations including the effects of the
dissipation. Some conclusions are given in
Sec.\ref{sec:conclusions}.

\section{ The model}
\label{sec:model} 
\begin{figure}
\centering
\includegraphics[scale=0.435]{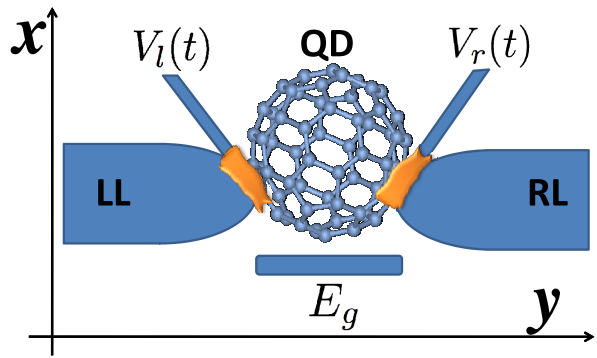}\\
\caption{The deformable quantum dot-based device consisting of a
molecular conductor connected to the external leads via two tunnel
barrier whose strength can be tuned periodically in time by using
suitable top gates. In the middle region an electric field $E_g$
affects the center of mass dynamics of the QD. Such dynamics takes
place in the direction perpendicular to the electrons transport.
The periodic out-of-phase raising/lowering of the barriers
strength enables the charge/discharge of the central region which
reacts to the electric field $E_g$ due to the electrostatic
force.} \label{fig:fig1}
\end{figure}

The system under study is shown in
Fig.\ref{fig:fig1}. It consists of a deformable quantum dot (QD)
coupled to two external leads, namely the right lead (RL) and the
left lead (LL), kept at the same chemical potential $\mu$. The
middle region is biased by a back gate which generates a
time-independent electric field $E_g$ that exerts an electrostatic
force $-eE_e n$, when the electron charge on the QD region is $-e
n$ (being  $-e$ the electron charge). The QD region, which can be
made of a nanotube or a $C_{60}$ molecule, is thus deformed under
the effect of $E_g$ and reaches its equilibrium configuration when
the elastic force is balanced. The electron transport in the QD is
generated by the adiabatic parametric pumping consisting in the
out-of-phase modulation of the two tunnel barrier strength
$V_{l,r}(t)$ at the interface with the external leads. \\
In order to study the transport through the quantum dot we
consider as classical the variables associated to the mechanical
deformation of the QD region, namely the position $x$ and the
momentum $P$ of the center of mass, while the electron transport
is treated within the quantum theory. Such approximation is
justified when the center of mass dynamics is much slower compared
to the electron dynamics (i.e. $\Omega \ll \Gamma/\hbar$, being
$\Omega$ the mechanical frequency of the dot and $\Gamma$ the
tunneling rate controlling the dot-leads dynamics). In the
following we consider a mechanical frequency $\Omega$ of the
central region comparable to the frequency $\omega$ of the pump
and thus of the order of $100-300$ MHz or less. The Hamiltonian
describing the system is the following:
\begin{eqnarray}
\label{eq:hamiltonian}
H&=&\sum_{\alpha=k,L,R}\epsilon_k c^{\dag}_{k\alpha}c_{k\alpha}+(\epsilon_0-eE_g x )d^{\dag}d\\\nonumber
&+&\sum_{k\alpha}(V_{\alpha}(t)d^{\dag}c_{k\alpha}+H.c.)\\\nonumber
&+&\frac{P^2}{2M}+\frac{M\Omega^2x^2}{2},
\end{eqnarray}
where the first term describes the Hamiltonian of the free
electrons in the LL and RL written in term of the creation
(annihilation) operators $c^{\dag}_{k\alpha}$ ($c_{k\alpha}$), the
second term describes the QD energy in terms of the displacement
$x$, the third term represents the tunneling between the leads and
the QD, while the last two terms describe the classical degrees of
freedom related to the center of mass motion of the QD region
having mass $M$ and bare resonance frequency $\Omega$. The energy
on the QD thus depends on the classical state of the oscillator
while the oscillator is affected by the charge state of the QD.\\
Differently from the charge shuttle case, where the tunneling
strength $V_{\alpha}$ depends on the longitudinal position of the
shuttle, in the present analysis such dependence is not important
and will be neglected in the following analysis. \\
From the
classical Hamilton equations, $\dot{P}=-\partial_x \langle H
\rangle $, $\dot{x}=\partial_P \langle H \rangle $ the following
equation for the center of mass motion\cite{nota2} is obtained:
\begin{equation}
\label{eq:dynamics-x}
M \ddot{x}+\beta \dot{x}+M\Omega^2 x=eE_g \langle d^{\dag}d \rangle.
\end{equation}
where $\beta \dot{x}$ represents a phenomenological dissipative
term. On the r.h.s. the term $eE_g \langle d^{\dag}d \rangle$ is
the electrostatic force acting on the deformable dot when the
charge density $-e\langle d^{\dag}d \rangle$ is present on it.
Under the adiabatic hypothesis, the  variable $x$ and the ac
pumping terms $V_{\alpha}(t)$ can be considered as slow varying
parameters and thus we can compute the retarded Green's function
of the quantum dot within the average-time
approximation\cite{fazio_prl_05}:
\begin{eqnarray}
\label{eq:gf}
G^r(E,t)&=&\frac{1}{E-\epsilon(x)+i\frac{\Gamma(t)}{2}}\\\nonumber
&&\epsilon(x)=\epsilon_0-eE_g x,
\end{eqnarray}
where the wide band limit for the self-energy has been considered,
while $\Gamma(t)=\sum_{\alpha}2 \pi \rho
V^{\ast}_{\alpha}(t)V_{\alpha}(t)$, $\rho$ being the density of
states of the external leads ($\rho_l=\rho_r=\rho$). The
modulation of the tunnel strengths $V_{\alpha}(t)$ is responsible
for an adiabatic modulation of the linewidth
$\Gamma(t)=\sum_{\alpha}\Gamma_{\alpha}(t)$, where
$\Gamma_{\alpha}(t)=\Gamma_0^{\alpha}+\Gamma_{\alpha}^{\omega}\sin(\omega
t+\varphi_{\alpha})$. In addition, under no bias voltage applied
to the external leads and within the zero temperature limit
($T=0$), the charge density $-e\langle d^{\dag}d \rangle$ on the
quantum dot region can be determined by the relation $\langle
d^{\dag}d \rangle=\pi^{-1}\int^{\mu}_{-\infty}Im\{G^r(E,t)\}dE$
whose explicit expression is given by:
\begin{equation}
\label{eq:charge-density}
\langle d^{\dag}d \rangle=\frac{1}{2}+\frac{1}{\pi}\arctan\Bigl(\frac{\mu-\epsilon(x)}{\Gamma(t)/2}\Bigl),
\end{equation}
$\mu$ being the chemical potential of the external leads. As shown
in (\ref{eq:charge-density}), the charge density on the dot
follows instantaneously the $x$-dynamics and thus the equation for
the mechanical degree of freedom can be solved at each time. Once
the solution for $x \equiv x(t)$ has been obtained, the retarded
Green's function of the dot can be calculated. \\
The instantaneous
current pumped in the lead $\alpha$ can be computed within the
Green's function technique by considering the variation of the
charge $q=-e \langle d^{\dag}d \rangle$  with respect to the small
and slow harmonic variations of the parameters
$X_i(t)$[\onlinecite{brouwer98}]: $\delta
q(t)=\sum_i\partial_{X_i}q\delta X_i(t)$. By setting
$X_1=\epsilon(x(t))$ and $X_2=\Gamma(t)$ and by making use of
Eq.(\ref{eq:charge-density}), the time derivative of $q(t)$ is
given by:
\begin{eqnarray}
\label{eq:charge-der}
\frac{dq}{dt}&=&-\frac{e}{2\pi}|G^{r}(E=\mu,t)|^2\{\Gamma(t)\partial_t\epsilon(x)\\\nonumber
&&+[\mu-\epsilon(x)]\partial_t \Gamma(t)\}.
\end{eqnarray}
 Noticing that
$\Gamma(t)=\sum_{\alpha}\Gamma_{\alpha}(t)$, the r.h.s. of
Eq.(\ref{eq:charge-der}), can be written as
$\sum_{\alpha}I^{\alpha}(\mu,t)$ and thus the expression for the
current pumped in the lead $\alpha$ within the zero temperature
limit is recovered\cite{fazio_prl_05}. Under time averaging over
one pumping period, the first term in Eq.(\ref{eq:charge-der})
vanishes and the relation $\bar{I}^{l}=-\bar{I}^{r}$ is verified.
The dc current pumped per cycle is given by:
\begin{equation}
\label{eq:dc-current}
\bar{I}^{\alpha}=\frac{\omega}{2\pi}\int_0^{2\pi/\omega}I^{\alpha}(\mu,t)dt,
\end{equation}
$\omega$ being the pumping frequency. In the following we focus on the weak pumping limit and confine
our analysis to the case of small deformation of the central
region. Within this approximation the equation of motion of the
$x$-variable can be linearized and the related solution can be
expressed in a  simple form.
\subsection{Dynamics of the mechanical degrees of freedom}
\label{sec:x-dyn}

In this subsection we treat the dynamics of the $x$-variable
within the linear approximation and consider the weak pumping
limit, $\Gamma^\omega_\alpha \ll \Gamma_0$. Let us first note that
$\langle d^{\dag}d\rangle$ depends on the instantaneous position
$x$ of the deformable quantum dot and such dependence induces a
non linear term in the dynamics as shown by
Eqs.(\ref{eq:dynamics-x}) and (\ref{eq:charge-density}). To
linearize Eq.(\ref{eq:dynamics-x}), we expand $x(t)$ around the
equilibrium position $\bar{x}$ and write $x(t)=\bar{x}+\xi(t)$.
Substituting $x(t)=\bar{x}+\xi(t)$ in Eq.(\ref{eq:charge-density})
we get the following weak pumping and small deformation expansion:
\begin{eqnarray}
\label{eq:charge-density-expansion}
\langle d^{\dag}d\rangle &\simeq& n_0-\frac{\lambda \xi(t)}{\pi}Im\{G_0^r(E=\mu)\}\\\nonumber
&&-\sum_{\alpha}\frac{\Gamma^{\omega}_{\alpha}(t)}{2\pi}Re\{G_0^r(E=\mu)\},
\end{eqnarray}
where $n_0$ is defined from Eq.(\ref{eq:charge-density}) by
setting $\xi(t)=0$ and $\Gamma^{\omega}_{\alpha}(t)=0$, while
$G^r_0(E=\mu)=[\mu-\tilde{\epsilon}_0+i\sum_{\alpha}\Gamma_0^{\alpha}/2]^{-1}$,
$\tilde{\epsilon}_0=\epsilon_0-\lambda \bar{x}$,
$\Gamma_{\alpha}^{\omega}(t)=\Gamma_{\alpha}^{\omega}\sin(\omega
t+\varphi_{\alpha})$. Substituting
Eq.(\ref{eq:charge-density-expansion}) in Eq.(\ref{eq:dynamics-x})
and by introducing the adimensional displacement $
\eta(\tau)=\xi/x_0$ and time  $\tau=\Omega t$ we get the following
linear equation:
\begin{eqnarray}
\label{eq:dynamics-eta}
\ddot{\eta}+Q^{-1}\dot{\eta}+\alpha^{2}\eta=-\frac{\bar{\lambda}}{2\pi}\sum_{\alpha}\Gamma^{\omega}_{\alpha}(\tau)Re\{G^r_0(E=\mu)\},
\end{eqnarray}
where $Q$ represents the quality factor of the oscillator (notice
that $\beta=M\Omega/Q$), $\bar{\lambda}=\lambda/(M\Omega^2x_0)$,
while $\alpha^2=\tilde{\Omega}^2/\Omega^2$, $\tilde{\Omega}$ being
the renormalized oscillation frequency:
\begin{eqnarray}
\label{eq:dressed-frequency}
\tilde{\Omega}=\sqrt{\Omega^2+Im\{G^r_0(E=\mu)\}\lambda^2/(\pi M)}.
\end{eqnarray}
The solution of Eq.(\ref{eq:dynamics-eta}) can be obtained by the
classical response function $\chi(\tau)$ which is solution of the
equation $\hat{D} \chi(\tau)=\delta(\tau)$ where
$\hat{D}\equiv[\frac{d^2}{d\tau^2}+Q^{-1}\frac{d}{d\tau}+\alpha^2]$
is a linear differential operator. Under the effect of the forcing
term $F(\tau)$, the r.h.s. of the equation
(\ref{eq:dynamics-eta}), the displacement $\eta(\tau)$ of the
deformable quantum dot can thus be obtained as
$\eta(\tau)=\int_{-\infty}^{\infty}d\tau'\chi(\tau-\tau')F(\tau')$.
Therefore the explicit solution for the $x$-variable is:
\begin{eqnarray}
\label{eq:x-solution}
x(t)&=&\bar{x}-\frac{|\chi|\bar{\lambda} x_0}{2\pi}Re\{G^r_0(E=\mu)\}\times\\\nonumber
&&\sum_{\alpha}\Gamma^{\omega}_{\alpha}\sin(\omega t-\phi_D+\varphi_{\alpha}),
\end{eqnarray}
where the quantities $|\chi|$  and $\phi_D$ come from the
exponential representation $|\chi|\exp(i\phi_D)$ of the Fourier
transform of the response function
$\chi(s)=(\alpha^2-s^2-isQ^{-1})^{-1}$,
\begin{equation}
|\chi(s)|=\frac{1}{\sqrt{(\alpha^2-s^2)^2+s^2Q^{-2}}}\nonumber
\end{equation}
and
\begin{equation}
\phi_D(s)=\arctan\Bigl(\frac{Q^{-1}s}{\alpha^2-s^2}\Bigl).\nonumber
\end{equation}
From Eq.(\ref{eq:x-solution}) is evident that the dynamical phase
$\phi_D$ of the classical response function $\chi$ plays an
important role in the quantum pumping. The knowledge of the
dynamics of the deformable quantum dot allows us to derive the
retarded Green's function Eq.(\ref{eq:gf}) and thus the pumped
current.
\subsection{ Pumped currents within the weak pumping case}
\label{sec:weak pump}

In this subsection we derive the expression of the charge current
adiabatically pumped under the influence of the center of mass
dynamics of the central region.  By introducing the retarded
Green's function in Eq.(\ref{eq:dc-current}) and performing an
expansion of the integrand up to second order terms in
$\Gamma_{\alpha}^{\omega}$, the current pumped in the left lead
$i_l$ in units of $-\frac{e\omega}{2\pi}$ can be written
as\cite{nota1}:
\begin{eqnarray}
\label{eq:current-bilinear}
i_l&=&i^l_{eq}-\frac{E_p |\chi|}{2\pi}|G_0^r(E=\mu)|^2Re\{G_0^r(E=\mu)\}\\\nonumber
&\times&Im\{G_0^r(E=\mu)\}\Gamma_l^{\omega}\Gamma_r^{\omega}[\sin(\phi_D)\cos(\varphi)(\Gamma_0^{r}-\Gamma_0^{l})\\\nonumber
&+&\sin(\phi_D)\Bigl(\frac{\Gamma_l^{\omega}}{\Gamma_{r}^{\omega}}\Gamma_0^{r}-\frac{\Gamma_r^{\omega}}{\Gamma_l^{\omega}}\Gamma_0^{l}\Bigl)-\cos(\phi_D)\sin(\varphi)(\Gamma_0^{l}+\Gamma_0^{r})],
\end{eqnarray}
where the \textit{polaronic energy} $E_p= \bar{\lambda}\lambda
x_0/2=\lambda^2/(2M\Omega^2)$ has been introduced. The current
$i^l_{eq}$ is the current pumped when the quantum dot is in its
equilibrium position $\bar{x}$ and whose expression is:
\begin{eqnarray}
i^l_{eq}=\frac{\Gamma_r^{\omega}\Gamma_l^{\omega}}{2}\sin(\varphi)Im\{G_0^r(E=\mu)\}Re\{G_0^r(E=\mu)\}.
\end{eqnarray}
Eq.(\ref{eq:current-bilinear}), which is the main result of this
work, clearly shows that the classical phase $\phi_D$, coming from
the $x$-dynamics, may strongly  affect  the current-phase relation
compared to the static situation where an odd behavior of the
current with respect to the pumping phase $\varphi$ is expected.
When $\lambda$ goes to zero, e.g. by switching off  the central
gate ($E_g=0$), the standard weak pumping current is recovered.
Furthermore, from detailed analysis of
Eq.(\ref{eq:current-bilinear}) one sees that when the pumping
frequency $\omega$ becomes very close to the renormalized
resonance frequency $\tilde{\Omega}$ the pumped current is
significantly modified  by the additional terms related to the
oscillation of the quantum dot. Even within the small displacement
approximation, the amplitude of the second term on the r.h.s. of
the Eq.(\ref{eq:current-bilinear}) proportional to $E_p Q$ can
become comparable to the term proportional to
$\Gamma_{l}^{\omega}\Gamma_{r}^{\omega}$ for suitable values of
$E_g$ and thus the system may present interesting
oscillation-induced effects on the pumped current. The
oscillation-related terms of the current consist of three distinct
contributions. Two of them are related to the pumping phase
$\varphi$ and are proportional to $\sin \phi_D$ and $\cos \phi_D$,
while the remaining one is independent from the pumping phase and
can be interpreted as a rectification term. Concerning the terms
containing the pumping phase, the current-phase relation is no
more an odd function of $\varphi$ due to the presence of terms of
the form $\sin(\phi_D)\cos(\varphi)$ and
$\cos(\phi_D)\sin(\varphi)$ that can be recognized as interference
terms. If the oscillator has a very high $Q$ (low dissipation),
the terms proportional to $\sin(\phi_D)$ become relevant only very
close to the resonance frequency (see Fig.(\ref{fig:fig2})), while
the term proportional to $\sin(\varphi)$ is the dominant one. Thus
for weak dissipation, the odd symmetry of the current-phase
relation is preserved almost everywhere in the parameters space.
On the other hand, when the quality factor of the oscillator is
decreased, i.e. for increasing values of the dissipation, the
additional terms with phase dependence different from the odd ones
start to become relevant. Our analysis thus relates the
dissipation of the center of mass motion to the presence of the
anomalous terms in the current-phase relation. \\
Let us note that the mechanism related to the presence of a
dynamical phase shift is more general than the context we are
describing here and can in principle be present every time a
quantum system is coupled to a subsystem with a classical dynamics
 ( as, for instance, a classical RLC circuit).\\
Finally, we also stress that the $\cos(\varphi)$ term of the
pumped current we derived in the present analysis has also been
obtained in different contexts. For example, in
 Ref.[\onlinecite{buttiker-inelastic-pump01}] where the effect of the inelastic scattering within a two terminal parametric pump
was treated by means of the \textit{fictitious leads method}. In
that work a third fictitious lead with a time dependent chemical
potential was introduced. In our case, even in the absence of
inelastic phenomena, the deformable quantum dot behaves similarly
to the fictitious lead of
Ref.[\onlinecite{buttiker-inelastic-pump01}] and thus a
$\cos(\varphi)$ term appears. Interference-like terms in the
current-phase relation were also obtained in
Ref.[\onlinecite{moskalets-buttiker04}] where an adiabatic quantum
pump in the presence of external ac voltages was considered.

\section{Results}
\label{sec:results}

\begin{figure}
\centering
\includegraphics[scale=0.9]{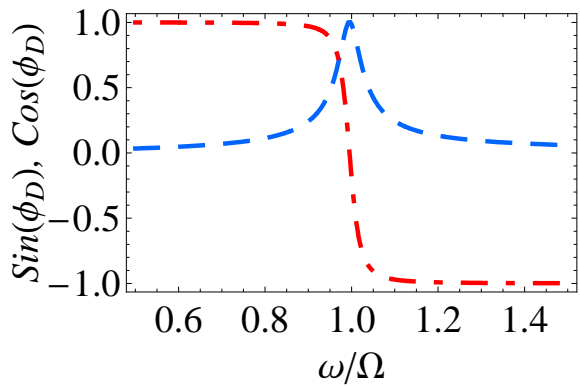}\\
\caption{Behavior of $\sin(\phi_D)$ (dashed line) and
$\cos(\phi_D)$ (dashed-dotted line) as a function of the
normalized pumping frequency $\omega$. The remaining parameters
have been fixed as follows: $\epsilon_0=0$, $E_p=0.015$,
$\Gamma_0^{l}=\Gamma_0^{r}=1, Q=20$. } \label{fig:fig2}
\end{figure}

In the following we fix the zero of the energy at the Fermi level
($\mu=0$) and consider the zero temperature limit to study the
pumped current (Eq.(\ref{eq:current-bilinear})) as a function of
the relevant system parameters. All the energies are expressed in
units of $\Lambda=10\mu$eV which is of the same order of magnitude
of the static linewidth $\Gamma^{\alpha}_0$ and the
\textit{polaronic energy} $E_p$ is assumed as a small quantity
compared to $\Lambda$. Since the \textit{polaronic energy} can be
tuned by means of the electric field $E_g$, in the following we
assume that the displacement of the central region is a fraction
of $10^{-12}$m and thus $E_p<10\mu$eV. In starting our analysis we
notice that the behavior of the pumped current depends strongly on
the value of the pumping frequency since both the response
function $\chi$ and the dynamical phase are functions of $\omega$.
This is clearly seen in Fig.(\ref{fig:fig2}) where $\sin(\phi_D)$
and $\cos(\phi_D)$ are shown as a function of the pumping
frequency $\omega$ (normalized to the mechanical frequency
$\Omega$) while the quality factor is fixed at a relatively low
value, $Q=20$ (high-dissipation).
When the mechanical dissipation is high, the $\sin(\phi_D)$
becomes a broad peak close to the renormalized mechanical
resonance, while $\cos(\phi_D)$ presents a step-like behavior as a
function of $\omega$ close to $\Omega$. Thus the presence of some
dissipation mechanism allows for the presence of an even function
of the pumping phase $\varphi$ in the current-phase relation.\\
\begin{figure}
\centering
\includegraphics[scale=0.9]{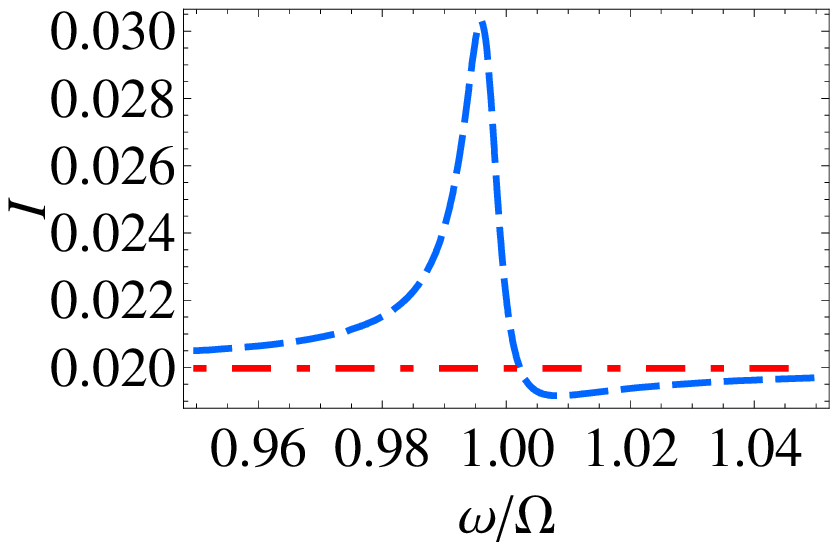}\\
\includegraphics[scale=0.925]{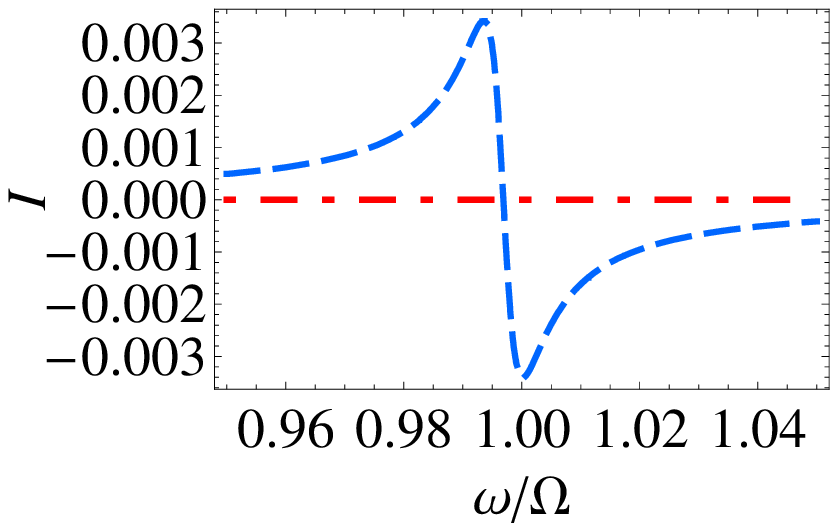}
\caption{Pumped current normalized to $-e\omega/(2\pi)$ computed
as a function of the pumping frequency $\omega$. The dashed-dotted
line corresponds to the current pumped in the absence of
\textit{polaronic} coupling (i.e. $E_p=0$), while the dashed
curves in both the panel represent the current pumped through the
system in the presence of mechanical oscillations, $E_p\neq 0$.
The upper panel is obtained for $\varphi=\pi/2$, while the lower
panel for $\varphi=0$. The remaining parameters have been fixed as
follows: $\epsilon_0=0.7$, $E_p=0.015$, $\Gamma_0^{l}=1.2$,
$\Gamma_0^{r}=1$, $\Gamma_{l}^{\omega}=0.3$,
$\Gamma_{r}^{\omega}=0.5$, $Q=150$.} \label{fig:fig3}
\end{figure}
In Figs.(\ref{fig:fig3}) we analyze the pumped current as a
function of the pumping frequency $\omega$ and by fixing the
remaining parameters as follows: $\epsilon_0=0.7$, $E_p=0.015$,
$\Gamma_0^{l}=1.2$, $\Gamma_0^{r}=1$, $\Gamma_{l}^{\omega}=0.3$,
$\Gamma_{r}^{\omega}=0.5$, $Q=150$, where we set $\varphi=\pi/2$
in the upper panel and $\varphi=0$ in the lower panel. The
dashed-dotted curves in both the panels represent the current
computed at $E_p=0$, i.e. in the absence of deformations  of the
central region, while the dashed line represents the current
pumped through the system when the system is coupled to the
classical phonons. Again we observe that when the system is driven
on resonance by the external ac parameters of the pump, the
oscillations of the dot are responsible for the enhancement of the
current close to the mechanical frequency $\Omega$, while for
pumping frequencies different from the resonance one the system
shows a current intensity similar to the static case. Furthermore,
the specific value of the pumping phase $\varphi$ can strongly
affect the symmetry of the pumped currents as a function of the
pumping frequency. Indeed, as is shown in the lower panel of
Figs.(\ref{fig:fig3}) for pumping frequencies around the resonance
frequency $\Omega$, the sign of the current is reversed.
\begin{figure}
\centering
\includegraphics[scale=0.9]{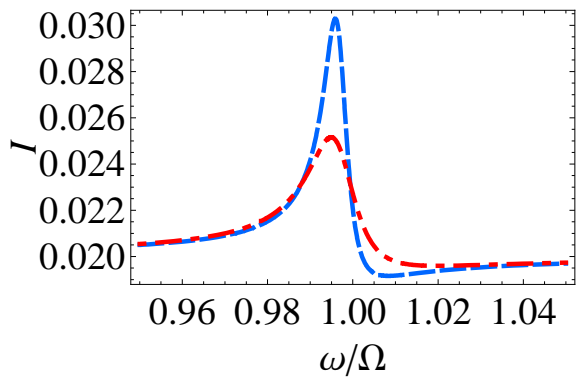}\\
\caption{Pumped current normalized to $-e\omega/(2\pi)$ computed
as a function of the pumping frequency $\omega$ and by fixing
$Q=150$ (dashed curve) and $Q=75$ (dashed-dotted curve). The
remaining parameters have been fixed as follows: $\varphi=\pi/2$,
$\epsilon_0=0.7$, $E_p=0.015$, $\Gamma_0^{l}=1.2$,
$\Gamma_0^{r}=1$, $\Gamma_{l}^{\omega}=0.3$,
$\Gamma_{r}^{\omega}=0.5$.} \label{fig:fig4}
\end{figure}
The behavior of the pumped current as a function of the pumping
frequency $\omega$ is qualitatively consistent with the one
measured in Ref.[\onlinecite{sazonova_nature04}] (see Fig.2a of
the cited work) where a suspended carbon nanotube with $Q=80$ and
$\tilde{\Omega}=55$MHz has been considered to analyze the
electrical-induced \textit{guitar-string-like } oscillation modes.
In Fig.(\ref{fig:fig4}) the pumped current is shown as a function
of the pumping frequency for two different values of the quality
factor, namely $Q=75$ (dashed-dotted line) and $Q=150$ (dashed
line), while the remaining parameters have been fixed as:
$\varphi=\pi/2$, $\epsilon_0=0.7$, $E_p=0.015$,
$\Gamma_0^{l}=1.2$, $\Gamma_0^{r}=1$, $\Gamma_{l}^{\omega}=0.3$,
$\Gamma_{r}^{\omega}=0.5$. As shown, higher values of $Q$ (lower
dissipation) permit to enhance the pumped current close to the
mechanical frequency.
\begin{figure}
\centering
\includegraphics[scale=0.9]{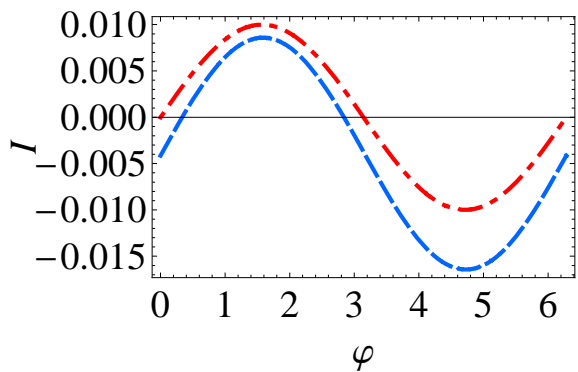}\\
\caption{Pumped current normalized to $-e\omega/(2\pi)$ computed
as a function of the pumping phase $\varphi$. The remaining
parameters have been fixed as follows: $Q=150$, $\omega/\Omega=1$,
$\epsilon_0=0.7$, $E_p=0.015$ (dashed line) or $E_p=0$
(dashed-dotted line), $\Gamma_0^{l}=1.2$, $\Gamma_0^{r}=1$,
$\Gamma_{l}^{\omega}=0.15$, $\Gamma_{r}^{\omega}=0.5$.}
\label{fig:fig5}
\end{figure}
The influence of the oscillation-related terms on the
current-phase relation is shown in Fig.(\ref{fig:fig5}) where the
pumped current is reported in the presence and not of the
deformations. As we see when $E_p\ne 0$ a nonzero pumped current
appears even for $\varphi=0, 2\pi$, i.e. a rectification term is
present.

\section{Conclusions}
\label{sec:conclusions} We studied the adiabatic quantum pumping
through a deformable quantum dot coupled to two external leads in
the presence of time varying barriers strength and analyzed the
oscillations induced effects on the pumped current. The
out-of-phase adiabatic modulation of the coupling to the leads is
responsible for the charging/discharging of the central deformable
region which is subject to an electrostatic field $E_g$. The
dynamics of the center of mass displacement $x$ of the central
region introduces an additional phase shift $\phi_D$ which affects
the pumped current. In particular the deformation induced terms
modify the current-phase relation by introducing even
contributions in the pumping phase $\varphi$. Such terms depend
strongly on the quality factor $Q$ and are suppressed  when the
mechanical dissipation is low. Furthermore, the current pumped is
enhanced when the pumping frequency $\omega$ is very close to the
mechanical resonance frequency $\tilde{\Omega}$, the enhancement
factor being determined by the $Q$ factor. Finally, we
demonstrated that the pumping mechanism can in principle be used
to characterize the mechanical properties of a molecular resonator
allowing to study the response function $\chi(t)$. Thus our
proposal can be seen as a complementary tool in the investigation
of the mechanical response of a nanoresonator. On the other hand,
the deformations of the central region permit to enhance the
pumped current for values of the pumping frequency close the to
mechanical one, thus rendering measurable the pumping current
which otherwise remains difficult to detect. Experimentally our
proposal can be realized by modifying the experimental set up
proposed in Ref.[\onlinecite{sazonova_nature04}] which is
particularly suitable due to the relatively low value of the
resonance frequency (i.e. $55$ MHz) fulfilling the adiabatic
requirement considered in our work.

\bibliographystyle{prsty}

\begin{thebibliography}{99}
\bibitem{nota1} The property $i_l=-i_r$ of the current is verified.
\end{thebibliography}

\end{document}